\documentclass[aps,prl, showpacs,twocolumn]{revtex4}
\usepackage{amssymb}
\usepackage{amsmath}
\usepackage{graphicx}
\usepackage{epsfig}

\begin{document}

\title{Wave Packet Transmission of Bloch Electron Manipulated by Magnetic
field}
\author{S. Yang$^{1}$, Z. Song$^{1,a}$ and C. P. Sun$^{1,2,a,b}$ }
\affiliation{$^{1}$Department of Physics, Nankai University, Tianjin 300071, China}
\affiliation{$^{2}$ Institute of Theoretical Physics, Chinese Academy of Sciences,
Beijing, 100080, China}

\begin{abstract}
We study the phenomenon of wave packet revivals of Bloch electrons and
explore how to control them by a magnetic field for quantum information
transfer. It is showed that the single electron system can be modulated into
a linear dispersion regime by the \textquotedblleft
quantized\textquotedblright\ flux and then an electronic wave packet with
the components localized in this regime can be transferred without
spreading. This feature can be utilized to perform the high-fidelity
transfer of quantum information encoded in the polarization of the spin.
Beyond the linear approximation, the re-localization and self-interference
occur as the novel phenomena of quantum coherence.
\end{abstract}

\maketitle

\section{Introduction}

Most recently, many theoretical investigations about quantum information
transfer (QIT) based on quantum spin systems are carried out in order to
implement scalable quantum computation \cite%
{Bose1,Ekert,ST,LY,SZ,key-1,TJO04,
key-17,key-6,key-18,BBG,PLENIO04,mpater}. Here, the quantum spin
system usually behaves as a quantum data bus to intergrade many
qubits. These investigations mainly aim at transferring quantum
state through a solid state data bus with minimal spatial and
dynamical control over the on-chip interactions between qubits. In
this paper we will pay attention to a fundamental aspect of QIT
and generally study the wave packet spreading and revival of Bloch
electrons in one-dimensional lattice systems.

For the problems of wave packet evolution, we can cast back for much earlier
investigations by Schr$\overset{..}{o}$dinger and others about the quantum
mechanical descriptions of localization of macroscopic objects \cite{PR1}.
They demonstrated that a class of wave packets (now we call them coherent
and squeezed states) of harmonic oscillator can keep their shapes during
propagation and their centers of mass (CM) follow a classical trajectory. As
a semi-classical solution of the Schr$\overset{..}{o}$dinger equation, a
superposition of the much higher excitation states with almost-homogeneous
spectrum form an coherent-state type wave packet in the Coulomb potential,
which can show the phenomena of non-spreading evolution and
self-interference on classical orbits \cite{PR2}. This prediction has been
demonstrated in the experiment involving the laser-induced excitation of
atomic Rydberg wave packets \cite{exp1, exp2}.

We can refer such non-spreading wave packet evolution with a complete
auto-correlation \cite{PR1} as a perfect QIT if we could encode the quantum
information in the spin polarization of electron. The investigation in this
paper is motivated by our recent explorations about the QIT based on the
quantum system possessing a commensurate structure of energy spectrum
matched with a symmetry (SMS), which ensures a perfect QIT both in one and
higher dimensional cases \cite{ST, LY2}. Actually the almost-homogeneous
spectrum for the coherent-state type wave packet just satisfies the
condition of SMS. In particular the non-spreading transfer of a
zero-momentum wave packet is attractive for the task of quantum information
transmission since a static superposition can behaves as a quantum storage.
This is very similar to the scheme of the quantum storage of photon based on
atomic ensemble where two stored photonic wave packets localized in the same
position with different polarizations can function to decode the information
of qubit \cite{prlsun,lukin}.

\begin{figure}[tbp]
\includegraphics[bb=77 380 512 640, width=6 cm, clip]{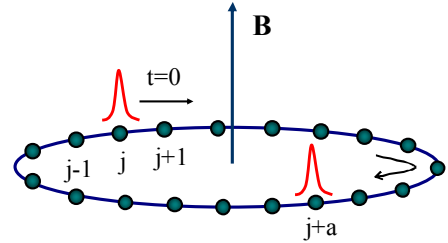}
\caption{\textit{(color online) The schematic illustration for the time
evolution of a wave packet in a ring threaded by a magnetic field.}}
\end{figure}

This paper will focus on a realistic, but simplest Bloch electron system
(see the Fig. 1) in a magnetic field where the on-site Coulomb interactions
are ignored. In this sense the spin polarization is always conversed during
the time evolution of an arbitrary state and then quantum information
encoded in the spin polarization of electron can be well protected. Thus the
locality of electron wave packet becomes a crucial element to maximize the
fidelity of QIT. We show that, by the \textquotedblleft
quantized\textquotedblright\ flux threading the ring lattice, the effective
dispersion relation of a Bloch electron can be modulated into a liner
dispersion regime that possesses SMS structure, and then an electronic wave
packet with the components localized in this linear regime can be
transferred without changes of its shape. This feature can be utilized to
perform the high-fidelity QIT encoded in the polarization of the spin. The
phenomena of wave packet revivals and self-interference can also be
demonstrated for the cases beyond the linear dispersion regime. These novel
quantum coherence effects may suggest a feasible protocol to implement the
perfect QIT of Bloch electrons manipulated by the external magnetic field.

\section{Model of flux-controlled Bloch electron in a ring and its
linearization}

In this section, we present the Bloch electron model under consideration, a
simple tight-binding model in an external magnetic field. Here, the Coulomb
interaction is ignored for simply. We restrict our attention to the
influence of the applied field on the propagation of the Bloch electrons.

Consider a ring lattice with $N$ sites threaded by a magnetic field
illustrated schematically in Fig. 1. The Hamiltonian of the corresponding
tight-binding model
\begin{equation}
H[\phi ]=-J\sum_{j,\sigma }(e^{i2\pi \phi /N}a_{j,\sigma }^{\dagger
}a_{j+1,\sigma }+h.c.),
\end{equation}%
depends on the magnetic flux $\phi $ through the ring in the unit of flux
quantum $\Phi _{0}=h/e$. Here, $a_{j,\sigma }^{\dagger }$ is the creation
operator of Bloch electron at $j$th site with spin $\sigma =\uparrow
,\downarrow $. The flux $\phi $ does not exert force on the Bloch electrons,
but can change the local phase of its wave function due to the Aharanov-Bohm
(AB) effect. Note that the interaction between the field and electrons is
independent of the intrinsic degree of freedom-spin. This will be crucial to
employ such kind of setup to transfer quantum information encoded in the
polarization of the spin. Because of the AB effect, the role of the magnetic
flux cannot be removed trivially.

Now we consider the evolution of the GWP%
\begin{equation}
\left\vert \psi _{\sigma }(k_{0},N_{A})\right\rangle =\frac{1}{\sqrt{\Omega
_{1}}}\sum_{j}e^{-\frac{^{\alpha ^{2}}}{2}(j-N_{A})^{2}}e^{ik_{0}j}\left%
\vert j\right\rangle _{\sigma }
\end{equation}%
with the velocity $k_{0}$, where $\left\vert j\right\rangle =a_{j,\sigma
}^{\dag }\left\vert 0\right\rangle $, the half-width of the wave packet
\begin{equation}
2\sqrt{\ln 2}/\alpha <<N,
\end{equation}%
and the normalization factor $\Omega _{1}=\sum_{j}e^{-\alpha
^{2}(j-N_{A})^{2}}$. The limitation for the width of the GWP ensures the
locality the state and avoids the overlap between the head and the tail of
the wave packet. We will see that as time evolution, the head and tail do
meet in certain situation and the interference phenomenon occurs.

\begin{figure}[tbp]
\includegraphics[bb=100 260 486 585, width=6 cm, clip]{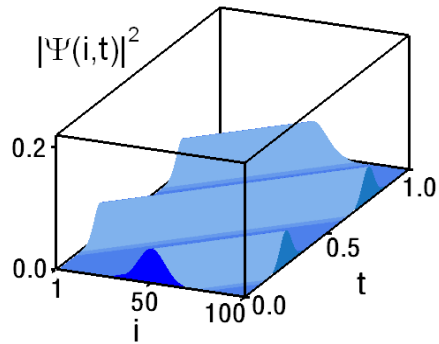}
\caption{\textit{(color online) Numerical simulation of the time evolution
of a zero-momentum Gaussian wave packet (GWP) with $\protect\alpha =0.1$ in
position-space $i$ of the 100-site ring with $\protect\phi =\protect\pi /2$.
The time $t$ is in the unit of $100/J$.}}
\end{figure}

In the following, we will show that the appropriate magnetic flux can ensure
the transfer of the wave packet without spreading. The well-known Bloch
dispersion relation
\begin{equation}
\varepsilon _{k}=-2J\cos (k+\frac{2\pi \phi }{N}).
\end{equation}%
can be obtained through the Fourier transformation
\begin{equation}
a_{k,\sigma }^{\dag }=\frac{1}{\sqrt{N}}\sum_{j}e^{ikj}a_{j,\sigma }^{\dag },
\end{equation}%
which can be employed to diagonalize $H[\phi ]$ as
\begin{equation}
H[\phi ]=\sum_{k,\sigma }\varepsilon _{k}n_{k,\sigma }.
\end{equation}%
with the number operator $n_{k,\sigma }=a_{k,\sigma }^{\dagger }a_{k,\sigma
} $.

It is observed that, when we tune the flux $\phi $ into each discrete values
\begin{equation}
\phi =\phi _{n},\equiv (\frac{1}{2}n+\frac{1}{4})N,
\end{equation}%
for $n=0,1,2,...,$there is a linear dispersion regime with momenta $k$
around zero, i.e., $\varepsilon _{k}\sim k$. For the wave packets as a
superposition of those eigenstates with momenta just in this region, then
the effective Hamiltonian becomes
\begin{equation}
H_{eff}=vp,
\end{equation}%
which is of the \textquotedblleft ultra-relativistic\textquotedblright\ type
with the effective \textquotedblleft light velocity\textquotedblright\ $%
v=(-1)^{n}2J$ and
\begin{equation}
p=\sum_{k,\sigma }ka_{k,\sigma }^{\dagger }a_{k,\sigma }
\end{equation}%
is the Bloch momentum operator. Obviously, this displacement effect time can
directly result in a non-spreading wave packets transmission in the linear
dispersion regime. Indeed. Fig. 3 illustrates how the \textquotedblleft
quantized\textquotedblright\ magnetic flux $\phi =\phi _{n}$ where $%
n=0,1,2,...,$ can speed up the zero-momentum Gaussian wave packet (GWP) $%
\left\vert \psi (0,N_{A})\right\rangle $ centered at site $N_{A}$ with the
width $1/\alpha $. The details of the will be given in the next section.

Actually, with the linearized Hamiltonian $H_{eff}$, the time evolution of
some states can be described as a spatial translation by the evolution
operator
\begin{equation}
U(t)=\exp (-ipvt)\equiv T(vt),
\end{equation}%
with a displacement $x=vt$. Here the the translational operator is defined
by
\begin{equation}
T(x_{0})\left\vert j\right\rangle _{\sigma }=\left\vert j+x_{0}\right\rangle
_{\sigma }
\end{equation}%
for arbitrary $\left\vert j\right\rangle _{\sigma }$.

For small $\alpha $, $\left\vert \psi (0,N_{A})\right\rangle _{\sigma }$ is
also a GWP of width $\alpha $ around $k=0$ in $k$-space, and then the wave
function at instance $t$ is a translated GWP
\begin{equation}
\left\vert \Phi (t)\right\rangle _{\sigma }=T(vt)\left\vert \psi
(0,N_{A})\right\rangle _{\sigma }=e^{i\varphi }\left\vert \psi
(0,N_{c}(t))\right\rangle _{\sigma },
\end{equation}%
which is centered at $N_{c}(t)=N_{A}+vt$. The overall phase factor $%
e^{i\varphi }$ has no effect on the final result. It is \ seen that the wave
packet moves with velocity $v$. Since all the wave functions satisfy the
periodic boundary condition
\begin{equation}
\left\vert \psi (0,N+i)\right\rangle _{\sigma }=\left\vert \psi
(0,i)\right\rangle _{\sigma },
\end{equation}%
we have
\begin{equation}
\left\{
\begin{tabular}{l}
$_{\sigma }\left\langle i\right. \left\vert \Phi (t)\right\rangle _{\sigma
}= $ $_{\sigma }\left\langle i-N\right. \left\vert \Phi (t)\right\rangle
_{\sigma }$ \ \ for $i>N,$ \\
$_{\sigma }\left\langle i\right. \left\vert \Phi (t)\right\rangle _{\sigma
}= $ $_{\sigma }\left\langle i+N\right. \left\vert \Phi (t)\right\rangle
_{\sigma }$ \ \ for $i<1.$%
\end{tabular}%
\right.
\end{equation}%
Obviously, the wave packet moves along the ring keeping the initial shape
without any spreading as illustrated in Fig. 2 schematically.

\section{Non-spreading wave packets evolution and solid state flying qubit}

To analyze the cyclic motion of Bloch electrons, the numerical simulation is
performed for a zero-momentum GWP with $\alpha =0.1$ in the system of $N=100$
and $\phi =N/4=25$. The simulated time evolution of wave packet is plotted
in Fig. 3. For the cases with different values of $\phi $ and $\alpha $, the
autocorrelation functions
\begin{equation}
\left\vert A(t)\right\vert =\sum_{\sigma }\left\vert _{\sigma }\left\langle
\Phi (t)\psi (0,N_{A})\right\rangle _{\sigma }\right\vert ,
\end{equation}%
which can be used to describe the properties of the electron propagation,
are investigated numerically. The results for $\phi =20,$ $25,$ $33$ and $%
\alpha =0.1,$ $0.3$ are plotted in Fig. 3(a) and 3(b), which show that a
zero-momentum GWP with small $\alpha $ can be transferred without spreading
when $\phi $ are around each $\phi _{n}$.

Meanwhile, the flux threading the ring can control the shape and destination
of the final wave packet. It is observed from the above analysis that the
flux plays an important role for manipulating non-spreading wave packet.
Actually, such phenomenon can also be understood by the following
transformation of the basis vector of Hilbert space. The existence of the
flux is equivalent to adding a speed to boost the zero-momentum wave packet
since the magnetic flux provides an extra phase to the basis in the
position-space, i.e.,
\begin{equation}
\left\vert j\right\rangle _{\sigma }\rightarrow \left\vert \psi (\phi
)\right\rangle _{\sigma }=e^{2\pi i\phi j/N}\left\vert j\right\rangle
_{\sigma }.
\end{equation}%
In other words, a GWP with small $\alpha $ and momentum $k_{0}=\pi (2n+1)/2$%
, can transfer along the ring without spreading approximately. We will
demonstrate this in the last section about open chain system.

\begin{figure}[tbp]
\includegraphics[bb=8 76 585 769, width=8 cm, clip]{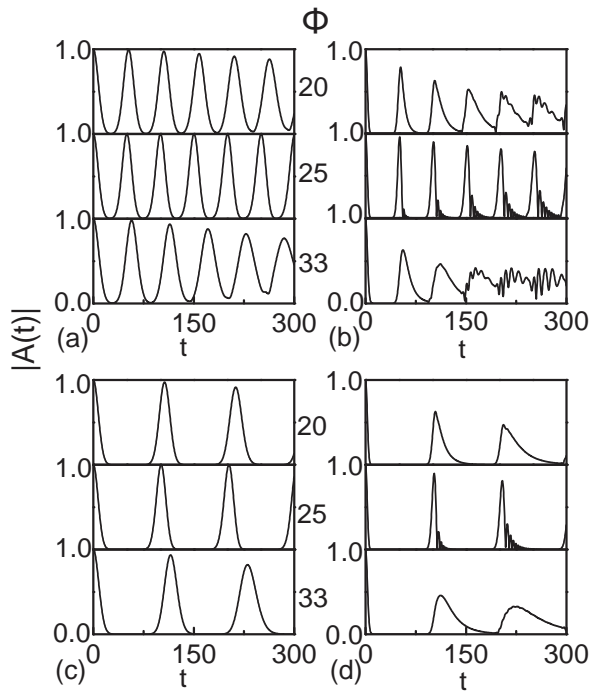}
\caption{\textit{Numerical simulation of the autocorrelation functions $%
|A(t)|$ of the zero-momentum GWPs (or $k_{0}=2\protect\pi \protect\phi /N$
for the system without external field) with $\protect\alpha =0.1$ (a, c) and
$\protect\alpha =0.3$ (b, d) in the 100-site ring (a, b) and chain (c, d)
with $\protect\phi =20,25,$ and $33$ (or $\protect\phi =0$ but the GWPS with
corresponding speed). It shows that for small $\protect\alpha $ and $2%
\protect\pi \protect\phi /N=\protect\pi /2$ the GWP can be transferred
without spreading. The unit of time $t$ is $1/J$.}}
\end{figure}

In the above studies, the spin state of the\ Bloch electron is a
conservative quantity that can not be influence during the propagation no
matter how the spatial shape of the wave function changes. From an abstract
point of view, the spatial properties of the carrying particle, i.e. the
Bloch electron, seems to be irrelevant since only amplitudes and relative
phases are used to encode quantum information. \ However, when the
propagation of the Bloch electron can be exploited to transfer the
information of qubit, the non-spreading propagation of the carrier is very
crucial for the expected high-fidelity of quantum state transfer from a
location to another.

With the above consideration we can imagine the electronic wave packets with
spin polarization as an analogue of photon \textquotedblleft flying
qubit\textquotedblright , the type II (polarized) photon qubit where the
quantum information was encoded in its two polarization states. We define
the solid state \textquotedblleft flying qubit\textquotedblright , at a
single location $A$ in a quantum wire, as the two Bloch electronic wave
packets $\left\vert 1\right\rangle _{A}=\left\vert 1\right\rangle \left\vert
N_{A}\right\rangle $ and $\left\vert 0\right\rangle _{A}=\left\vert
0\right\rangle \left\vert N_{A}\right\rangle $ be encoding as
\begin{eqnarray}
\left\vert 1\right\rangle \left\vert N_{A}\right\rangle  &=&\frac{1}{\sqrt{%
\Omega _{1}}}\sum_{j}e^{-\frac{^{\alpha ^{2}}}{2}(j-N_{A})^{2}}e^{i\frac{\pi
}{2}j}a_{j,\uparrow }^{\dag }\left\vert 0\right\rangle ,  \notag \\
\left\vert 0\right\rangle \left\vert N_{A}\right\rangle  &=&\frac{1}{\sqrt{%
\Omega _{1}}}\sum_{j}e^{-\frac{^{\alpha ^{2}}}{2}(j-N_{A})^{2}}e^{i\frac{\pi
}{2}j}a_{j,\downarrow }^{\dag }\left\vert 0\right\rangle .
\end{eqnarray}%
Because of the intrinsic linearity of the Schr$\overset{..}{o}$dinger
equation, it is self-consistent to encode an arbitrary state
\begin{equation}
\left\vert \psi \right\rangle =\cos (\frac{\theta }{2})\left\vert
1\right\rangle _{A}+\sin (\frac{\theta }{2})e^{i\varphi }\left\vert
0\right\rangle _{A}
\end{equation}%
as
\begin{equation}
\left\vert \psi (\theta ,\varphi )\right\rangle _{A}=\frac{1}{\sqrt{\Omega
_{1}}}\sum_{j}e^{-\frac{^{\alpha ^{2}}}{2}(j-N_{A})^{2}}e^{i\frac{\pi }{2}%
j}a_{j,\sigma }^{\dag }\left\vert 0\right\rangle
\end{equation}%
and then it is transferred to another place $B$ with a very high fidelity
due to the feature of non-spreading propagation of Bloch electron.

\section{Self-interference and revival of spreading wave packet}

We now turn our attention to the problem of non-spreading propagation of
Bloch electronic wave packets beyond the linear dispersion regime. We
consider a zero-momentum GWP in an external field with $\phi $ far from $%
\phi _{n}$ (or a GWP with small momentum $k_{0}$ but $\phi =0$). Because of
the nonlinear dispersion relation, such kind of wave packet spreads while
its center is moving. It is clear that, when the head of the wave packet
catches up with its tail, quantum interference phenomena set in.

In order to demonstrate this phenomenon, numerical simulation is performed
for a GWP with $\alpha =0.3$ and $k_{0}=0.05\pi $ (or a zero-momentum GWP
with the external magnetic flux $\phi =k_{0}N/2\pi $) in the $100$-site ring
system. The time evolution of the GWP obtained by numerical simulation is
plotted In Fig. 4(a). The interference fringe appears when the GWP spreads,
which demonstrates the self-interference phenomenon. The profile of the
fringe can be estimated analytically as follows.
\begin{figure}[tbp]
\includegraphics[bb=52 182 549 640, width=8 cm, clip]{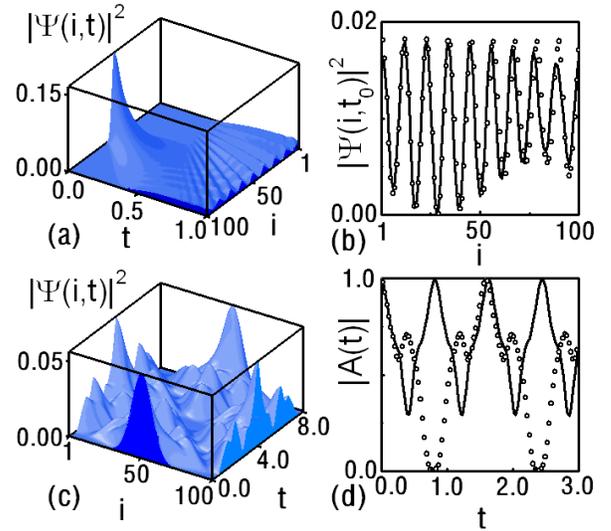}
\caption{\textit{(color online) The self-interference (a) and quantum
revival (c) phenomena of the GWPs obtained by numerical simulation. (b)
Plots of the self-interference fringe for 100-site ring obtained by
numerical simulation (solid line) and theoretical analysis (circle) at $%
t_{0}=\protect\delta \protect\tau =90/J$. (d) Plots of the autocorrelation
functions, $|A(t)|$, for the zero-momentum GWPs with $\protect\alpha =0.1$
in 100-site ring (circle) and chain (solid line) systems. The unit of time t
is 100/J in (a), (c), and 1000/J in (d).}}
\end{figure}

Consider a GWP $\left\vert \psi (k_{0},N_{A})\right\rangle $ at $t=0$ in the
coordinate space. When $\phi =0$, the Hamiltonian $H[0]$ can be
approximately written as
\begin{equation}
H_{eff}=-J\sum_{k,\sigma }\varepsilon _{k}a_{k,\sigma }^{\dagger
}a_{k,\sigma }
\end{equation}%
for $\varepsilon _{k}\sim (2-k^{2})$. On the other hand, $\left\vert \psi
(k_{0},N_{A})\right\rangle _{\sigma }$ is also a GWP around $k_{0}$ in the $%
k $-space. Then for GWP with small $k_{0}$ will evolves into a GWP%
\begin{equation}
\left\vert \Phi (t)\right\rangle _{\sigma }=A_{2}\sum_{j}e^{i\varphi
(j,t)}e^{-\frac{\alpha ^{\prime 2}}{2}(j-N_{c})^{2}}\left\vert
j\right\rangle _{\sigma }
\end{equation}%
with the spreading width
\begin{equation}
\alpha ^{\prime }=\alpha /\sqrt{1+4\alpha ^{4}J^{2}t^{2}},
\end{equation}%
centered at $N_{c}=N_{A}+2Jk_{0}t$, where $A_{2}$ is normalization factor,
and%
\begin{equation}
\varphi (j,t)=k_{0}j+2Jt-Jk_{0}^{2}t+Jt(j-N_{c})^{2}\alpha ^{2}\alpha
^{\prime 2}
\end{equation}%
is the time-dependent phase, i.e., the momentum of the moving GWP. Such kind
of wave packet spreads while its center is still moving. Since all the wave
functions satisfy the periodic boundary condition $\left\vert
N+j\right\rangle =\left\vert j\right\rangle $, at certain instant $t$, there
is an overlap between the head and tail parts of the wave packet and then
the quantum interference phenomena occurs. In other words, when the GWP
spreads over the circumference of the ring, we need to consider the virtual
superposition of the \textquotedblleft head\textquotedblright\ and
\textquotedblleft tail\textquotedblright .

Since the widely spreading GWP can wind the ring many times, the virtual
superposition can be considered as the re-normalized wave function
\begin{equation}
_{\sigma }\langle j\left\vert \Phi _{vs}(t)\right\rangle _{\sigma }=\frac{1}{%
\sqrt{\Omega _{vs}}}\text{ }_{\sigma }\langle j\pm lN\left\vert \Phi
(t)\right\rangle _{\sigma }
\end{equation}%
where
\begin{equation*}
\Omega _{vs}=\sum_{j}\left\vert _{\sigma }\langle j\left\vert \Phi
_{vs}(t)\right\rangle _{\sigma }\right\vert ^{2}
\end{equation*}%
is a normalization factor. For small $t=\delta \tau $, one can only take the
summation over $l=0,1$ as approximation, which results in spatial
interference fringe
\begin{eqnarray}
\left\vert _{\sigma }\langle j\left\vert \Phi _{vs}(\delta \tau
)\right\rangle _{\sigma }\right\vert ^{2} &=&\frac{1}{\sqrt{\Omega _{vs}}}%
\left\vert _{\sigma }\langle j\left\vert \Phi (\delta \tau )\right\rangle
_{\sigma }\right\vert ^{2}  \notag \\
&&\times \lbrack 1+c^{2}+2c\cos (Kj+\varphi _{0})]
\end{eqnarray}%
with the effective wave vector%
\begin{equation}
K=2NJ\alpha ^{2}\alpha ^{\prime 2}\delta \tau
\end{equation}%
and initial phase
\begin{equation}
\varphi _{0}=K(N/2-N_{c})+k_{0}N.
\end{equation}%
Here, $\left\vert _{\sigma }\langle j\left\vert \Phi (\delta \tau
)\right\rangle _{\sigma }\right\vert ^{2}$ and
\begin{equation}
c=\exp [-\alpha ^{\prime 2}N(j-N_{c}+N/2)]
\end{equation}%
only provide the modulation to the fringe. The spatial period%
\begin{equation}
\Delta =\left\vert \frac{2\pi }{K}\right\vert =\left\vert \frac{\pi }{%
NJ\alpha ^{2}\alpha ^{\prime 2}\delta \tau }\right\vert ,
\end{equation}%
characterizes the interference fringe.

In Fig. 4(b), the interference fringe at $\delta \tau =90/J$ obtained by
numerical simulation and the analytical approximate result are plotted. It
shows that the theoretical analysis is in agreement with the result of
numerical simulation.

Now we consider the special case of zero-momentum GWPs moving along a
lattice without magnetic field. In this case, although the dispersion
relation for such kind of GWPs is nonlinear, the quantum revival is still
possible since the $k^{2}$-dispersion also meets the condition of SMS \cite%
{ST, LY2}. To demonstrate this numerical simulation for the time evolution
of a GWP with $\alpha =0.1$ in a ring of $100$-sites is performed and the
density probability of the GWP as the function of the position $i$ and time $%
t$ (in the unit of $100/J$) are plotted in Fig. 4(c). It shows that revival
occurs after the GWP spreads. According to the theory of SMS, the revival
time is $\tau =2\pi /\Delta E$ in the general cases, where $\Delta E$ is the
greatest common divisor of energy-level spacing between any two eigen
states. Then in general case, we have%
\begin{equation}
\tau =\frac{2}{\pi J}(N+1)^{2}
\end{equation}%
for the chain while $\tau =N^{2}/(2\pi J)$\ for the ring. Now we consider a
special case, in which the initial zero-momentum GWP is centered at the
middle of the chain. Obviously, the parity of the GWP with respect to the
reflection symmetry is even. Thus expansion coefficients of the GWP for all
the eigen states with odd parity are all zero, which means that the
effective levels driven the GWP are only the half set. Thus this fact
results in a particular revival time%
\begin{equation}
\tau =\frac{1}{4\pi J}(N+1)^{2}.
\end{equation}

To verify the above analysis, the autocorrelation functions are also
calculated for the initial zero-momentum GWPs with $\alpha =0.1$ and $%
N_{A}=(N+1)/2$\ in the ring of $100$-sites and chain. The results are
plotted in Fig. 4(d), which show that the revival time is well in agreement
with the analytical estimation.

\begin{figure}[tbp]
\includegraphics[bb=109 188 486 659, width=6 cm, clip]{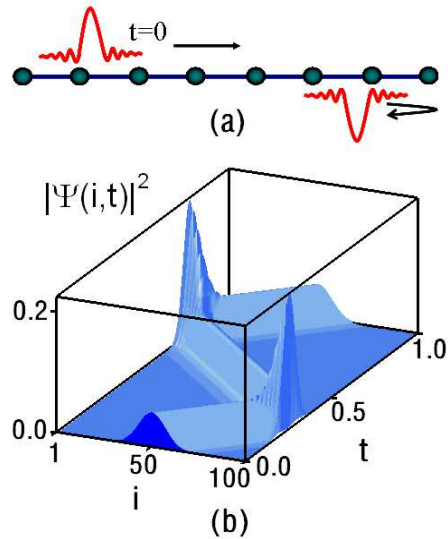}
\caption{\textit{(color online) (a) The schematic illustration for the time
evolution of a GWP in a chain. The $\protect\pi -shift$ occurs at the
boundary. (b) Numerical simulation of the time evolution of a moving GWP
with $\protect\alpha =0.1$ and $k_{0}=\protect\phi =\protect\pi /2$ in the
100-site chain. The unit of time t is 100/J.}}
\end{figure}

\section{The wave packet dynamics in the open chain}

In usual, a practical Bloch electron system is of open chain, in which the
magnetic filed has no longer effect on shape of the GWP. Corresponding to
the single-particle spectrum
\begin{equation}
\varepsilon _{k}=-v\cos k
\end{equation}%
where $v=2J$, the eigenvectors of $H$ are
\begin{equation}
\left\vert \psi _{k,\sigma }\right\rangle =a_{k,\sigma }^{\dag }\left\vert
0\right\rangle =\sum_{i=1}^{N}\sqrt{\frac{2}{N+1}}\sin (ki)\left\vert
i\right\rangle _{\sigma },
\end{equation}%
where%
\begin{equation}
k=\frac{\pi l}{N+1}
\end{equation}%
for $l=1,...,N$ can be regarded as a pseudo-momentum. Nevertheless, The GWP $%
\left\vert \psi _{\sigma }(k_{0},N_{A})\right\rangle $ located at $N_{A}$
with momentum $k_{0}$ $\sim \pi /2$ at $t=0$ will also evolve into
\begin{equation}
\left\vert \Phi _{k,\sigma }(t)\right\rangle =\left\vert \psi _{k,\sigma
}(k_{0},N_{A}+vt)\right\rangle .
\end{equation}%
Since all the eigenvectors satisfy the open boundary condition, we have
\begin{equation}
\left\vert \psi _{k,\sigma }(k_{0},N_{A}-vt)\right\rangle =-\left\vert \psi
_{k,\sigma }(k_{0},2N+2-N_{A}+vt)\right\rangle
\end{equation}%
for $N_{A}-vt>N$. It indicates that the wave packet reflects at the
boundaries with \textquotedblleft $\pi $-phase shift\textquotedblright .
Then the wave packet bounds back and forth along the chain as illustrated in
Fig. 5(a) schematically.

Similarly, under the transformation
\begin{equation}
e^{ik_{0}j}\left\vert j\right\rangle _{\sigma }\rightarrow \left\vert
j\right\rangle _{\sigma },
\end{equation}%
the propagation of a moving GWP with $k_{0}=2\pi \phi /N$ is equivalent to
that of a zero-momentum GWP in the system with extra phase $\exp (i2\pi \phi
/N)$ on the hopping term. Numerical simulation for the time evolution of a
GWP with $\alpha =0.1$ and $k_{0}=\pi /2$ in a chain of $N=100$ is plotted
in Fig. 5(b). The autocorrelation functions $\left\vert A(t)\right\vert $
are also calculated for $\alpha =0.1$, $0.3$ and $k_{0}=2\pi \phi /N$, $\phi
=20$, $25$, $33$, which are plotted in Fig. 3(c) and 3(d). It shows that a
GWP with small $\alpha $ and momentum $k_{0}=\pi (2n+1)/2$, can be
transferred along the chain without spreading approximately. From the
autocorrelation functions for rings and chains, it is easy to find that the
period of the revivals of GWP in a ring is approximately the half of that in
a chain. This is in agreement with the analytical results that the period is
$\tau =(N+1)/J$ for the chain and $\tau =N/(2J)$ for the ring. Comparing the
revival times for GWPs with and the linear and nonlinear dispersion
relation, we have the conclusion that the former is suitable for
implementing the fast QIT in the solid.

\section{Summary}

In summary, the quantum transmission of a Bloch electron in the
one-dimensional lattice is studied by theoretical analysis and numerical
simulation. It is found that a zero-momentum GWP can be transferred without
spreading approximately if an optimal magnetic flux is applied. This feature
can be employed to perform the high-fidelity QIT encoded in the polarization
of the Bloch electrons. Meanwhile, beyond such optimal range of the field,
the time evolution of the GWP is also investigated in the non-liner
dispersion regime. The novel quantum coherence effects found in this paper,
such as the wave packet revivals and self-interference, can motivate a
feasible protocol based on the practical systems to implement the perfect
QIT of Bloch electrons controlled by the external magnetic field.

We acknowledge the support of the CNSF (grant Nos. 90203018, 10474104) and
the National Fundamental Research Program of China (No. 001GB309310). One
(CPS) of the authors thanks the useful discussions with T. Xiang and Z.B. Su

\end{document}